\documentclass{INTERSPEECH2023}
\interspeechcameraready

\usepackage{caption}
\usepackage{stfloats}
\usepackage{multirow}
\usepackage{times}
\usepackage{epsfig}
\usepackage{amsmath}
\usepackage{amssymb}
\usepackage[ruled,linesnumbered]{algorithm2e}
\usepackage{graphicx}
\usepackage{algorithmic}
\usepackage{threeparttable}
\usepackage{subfigure}
\usepackage{ifpdf}

\usepackage{booktabs}
\usepackage{threeparttable}

\usepackage{multicol}
\usepackage{multirow}

\usepackage{color}

\usepackage{makecell} %^ 使用<{\centering}需引入宏包\usepackage{makecell}。

\usepackage[hang,flushmargin]{footmisc} % 在Latex中，使注脚首行不缩进，且新行与首行对齐，只需加这个宏包即可

\title{Learning Local to Global Feature Aggregation for Speech Emotion Recognition}
\name{Cheng Lu$^{\dag,1,2}$, Hailun Lian$^{\dag\thanks{$^\dag$These authors are contributed equally to this work.},1,3}$, Wenming Zheng$^{*\thanks{$^*$Corresponding Authors.},1,2}$, Yuan Zong$^{*,1,2}$, Yan Zhao$^{1,3}$, Sunan Li$^{1,3}$}
\address{
  $^1$Key Laboratory of Child Development and Learning Science of Ministry of Education, \\ Southeast University, Nanjing, China\\
  $^2$School of Biological Science and Medical Engineering, Southeast University, Nanjing, China \\
  $^3$School of Information Science and Engineering, Southeast University, Nanjing, China}
\email{\{cheng.lu, lianhailun, wenming\_zheng, xhzongyuan\}@seu.edu.cn}

\begin{document}

\maketitle

\begin{abstract}
% 1000 characters. ASCII characters only. No citations.
Transformer has emerged in speech emotion recognition (SER) at present. However, its equal patch division not only damages frequency information but also ignores local emotion correlations across frames, which are key cues to represent emotion. To handle the issue, we propose a Local to Global Feature Aggregation learning (LGFA) for SER, which can aggregate long-term emotion correlations at different scales both inside frames and segments with entire frequency information to enhance the emotion discrimination of utterance-level speech features. For this purpose, we nest a Frame Transformer inside a Segment Transformer. Firstly, Frame Transformer is designed to excavate local emotion correlations between frames for frame embeddings. Then, the frame embeddings and their corresponding segment features are aggregated as different-level complements to be fed into Segment Transformer for learning utterance-level global emotion features. Experimental results show that the performance of LGFA is superior to the state-of-the-art methods.

%Transformer has emerged in speech processing at present. However, its equal patch division not only damages the frequency information but also ignores local emotional correlations across frames for speech emotion recognition (SER), which are important cues to represent speech emotions. To cope with the issues, we propose a novel Local to Global Feature Aggregation learning (LGFA) method for SER. The advantage of LGFA is that it aggregates long-term emotion-related dependencies at different scales both inside frames and segments with entire frequency information to enhance the emotion discrimination of utterance-level speech features. For this purpose, we nest a Frame Transformer inside a Segment Transformer. Firstly, the Frame Transformer is design to excavate local emotion correlations between frames for the frame embeddings. Then, both frame embeddings and those corresponding segment features are aggregated as the complementary emotion correlations at different levels and fed into the Segment Transformer to learn the utterance-level global emotion features. The whole joint learning of two Transformers is from frame-level to segment-level to utterance-level. The experimental results demonstrate that our proposed LGFA improves the accuracies on IEMOCAP (Scripted$+$Improvised: $\uparrow6.25\%$ on WAR and $\uparrow7.82\%$ on UAR; Improvised: $\uparrow0.95\%$ on WAR) and CASIA ($\uparrow3.17\%$ on WAR and UAR) compared with the state-of-the-art methods.
\end{abstract}

\section{Introduction}

Speech emotion recognition (SER) is a significant task of affective computing and has attracted wide attention in recent years \cite{schuller2021intelligent}, \cite{cowie2001emotion}. The key to addressing the SER is how to disentangle the emotion information hidden in speech from the confusion of diverse acoustic factors \cite{schuller2009acoustic}, \cite{stuhlsatz2011deep}, \cite{lu2022domain}, e.\,g., background noise, language, speaker identity.

Actually, the emotional information is always discretely distributed in frames or segments of speech \cite{chen20183}, \cite{zhang2008design} due to the presence of special frames or segments without emotional contexts, i\,.e., empty frames/segments. In other words, emotion information is always discretely distributed in some key frames or segments. Therefore, a practical approach is to capture long-range emotion dependencies from these key frames/segments \cite{schuller2009acoustic}, \cite{stuhlsatz2011deep}, \cite{satt2017efficient}, \cite{wu2019speech}. To this end, Recurrent Neural Networks (RNNs) \cite{satt2017efficient}, \cite{wu2019speech} are widely adopted for learning utterance-level emotion features from frame-level or segment-level features.

Although previous works based on RNNs, e.\,g., LSTM and Bi-LSTM, have achieved great success on SER, they still encounter some issues \cite{yu2019review}, e.\,g., high time and space complexity for computing cells and only modeling sequential long-term dependencies (from forward to backward, or reverse). With the emergence of Transformer \cite{vaswani2017attention}, these issues have been handled effectively. In Transformer, the Multihead Self-Attention can describe the complete relationship between all speech frames/segments. Also, the time-space complexity could be effectively reduced by the matrix parallel calculation. Taking these advantages, the Speech Transformer models \cite{tarantino2019self}, \cite{gong2021ast} are promisingly developed from the Vision Transformer (ViT) \cite{dosovitskiy2020image}.

% \footnote{To disambiguate the usage of “patch” in speech processing and computer vision, we use “chunk” instead of “patch” in speech Transformer for SER.}
However, Speech Transformer roughly divides the speech spectrogram into same "chunks" \cite{lin2021chunk} (i.\,e., patches in ViT), leading to lossing local inter-frame relationships reflecting the fine-gained emotion distribution and corruption of frequency domain information. Since the frame-level and segment-level features contain the emotional information at different scales \cite{zhang2017speech}, \cite{shen2020wise}, e.g., frames reflect the phoneme-level associations and segments respond to the word-level or phrase-level correlations, they should be aggregated complementarily to learn more emotion-discriminative speech features. Likewise, ViT also ignores the local structure information in image patches for computer vision. To handle the similar issue, Han et al. \cite{han2021transformer} proposed a Transformer in Transformer (TNT) to simultaneously learn inter-patch and intra-patch relationships.

%TNT focuses on computer vision and uses equally divided small patches in the spatial domain of images as the input of inner Transformer. In SER, however, due to the difference in the time-frequency resolution on the spectrogram (usually as an input feature for the deep network), equally dividing the spectrogram in the time and frequency domains will damage the emotional correlation between frames or frequency bands. Thus, in this paper, we keep the information of the entire frequency domain and only divide patches in the time domain.

Inspired by TNT \cite{han2021transformer}, we propose a novel Local to Global Feature Aggregation learning (LGFA) method for SER. The LGFA nests a Frame Transformer inside a Segment Transformer to aggregate different-scale emotion dependencies for the speech emotion representation. The whole learning processing of LGFA is from frame-level to segment-level to utterance-level. Compared with other Speech Transformer-based methods, our LGFA is a novel and special Transformer-based model for SER and its advantages can be summarized as the following three folds:
\begin{enumerate}
  \item \emph{it aims to capture long-range emotion-related dependencies at different scales both inside frames and segments instead of the simple image patches adopted in Transformer.}
  \item \emph{it takes a frame and a segment as the input of Frame Transformer and Segment Transformer, respectively, instead of equally divided image patches.} In this case, the frame and segment used in LGFA may contain the entire frequency domain information such that the frequency feature will not be damaged in the speech chunk division.
  \item \emph{it also can be extended from the time domain to the frequency domain and time-frequency domain by different patch partition strategies.} This extension can make full use of the time-frequency characteristic of speech signals to represent emotion information.
\end{enumerate}

\section{Proposed Method}

\begin{figure}[t]
\centering
\includegraphics[width=3.3in,height=2.7in]{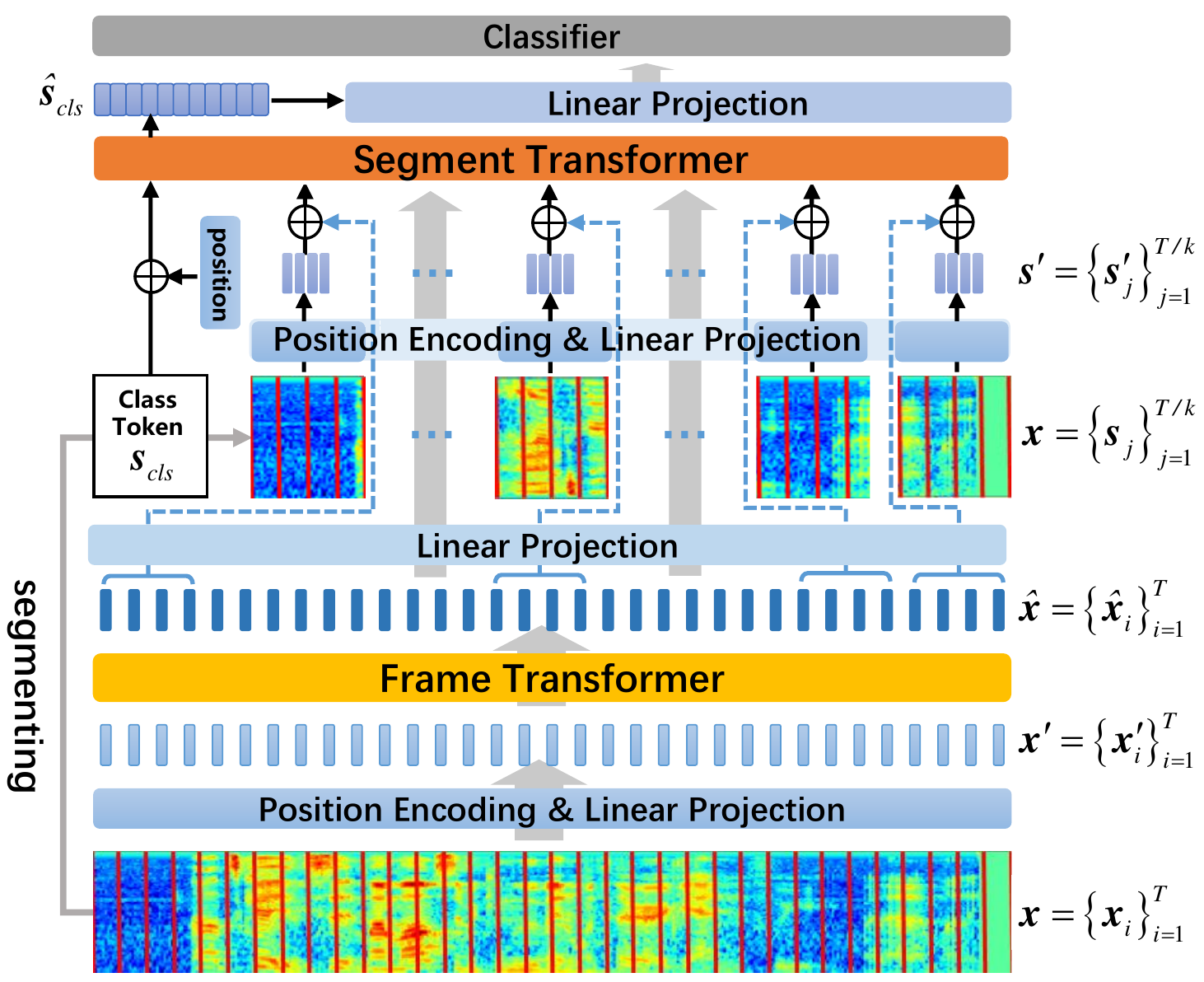}
\caption{Overview of Local to Global Feature Aggregation (LGFA) framework for SER. LGFA includes two Transformer-based networks, in which a Frame Transformer is nested inside a Segment Transformer.}
\label{fig:framework}
\end{figure}

Considering the inter-frame time property of speech, LGFA feeds a Frame Transformer with frame features, then integrates frame embeddings and segment features as the segment-level aggregation features. This point is the main difference from TNT. Further, these aggregation features are regarded as the input of a Segment Transformer to learn higher-level emotion correlations across segments. Consequently, we can obtain the global utterance-level features of speech emotions through joint training of the Frame and Segment Transformers. The overview of LGFA is shown in Figure. \ref{fig:framework}, in which the Frame Transformer takes the frame-level feature of speech as the input.

To this end, we firstly process the frame-level feature of speech. Given the log-Mel-spectrogram feature $\boldsymbol{x}\in\mathbb{R}^{F \times T \times C}$ of each emotional speech, the $i^{th}$ frame $\boldsymbol{x}_i\in\mathbb{R}^{F \times C}$ of the spectrogram $\boldsymbol{x}=\{\boldsymbol{x}_i\}_{i=1}^{T}$ is firstly encoded by a linear projection layer $\emph{FC}(\cdot)$ as the $i^{th}$ frame embedding $\boldsymbol{x}'_i \in \mathbb{R}^{1 \times d_f}$, denoted as
\begin{equation}
\boldsymbol{x}'_i = \emph{FC}(\boldsymbol{x}_i),
\label{eq:fc}
\end{equation}
where $F$, $T$, and $C$ represent the numbers of Mel-scaled frequency, time frame and channel, respectively. $d_f$ is the dimension of frame embeddings. Then, to enhance inductive bias of Frame Transformer \cite{dosovitskiy2020image}, we add a learnable position encoding $\boldsymbol{e}^f_i \in \mathbb{R}^{1 \times d_f}$ into $\boldsymbol{x}'_i$ as the input of Frame Transformer, which can be represented as
\begin{equation}
\boldsymbol{x}'_i \leftarrow \boldsymbol{x}'_i + \boldsymbol{e}^{f}_{i},
\label{eq:frame-po}
\end{equation}
where $\boldsymbol{e}^{f}=\{\boldsymbol{e}^f_i\}_{i=1}^T \in \mathbb{R}^{T \times d_f}$. In Frame Transformer, the sequence of speech frame embeddings $\boldsymbol{x}'=\{\boldsymbol{x}'_i\}_{i=1}^{T}$ is utilized to characterize local inter-frame correlations of emotions. Then, the frame-level encoding $\boldsymbol{\hat{x}}$ can be obtained by the frame embedding sequence $\boldsymbol{x}'$ through the following operations:
\begin{equation}
\label{eq:inner1}
\boldsymbol{x}''^{,\ell} = \emph{MSA}(\emph{LN}(\boldsymbol{x}'^{,\ell-1})) + \boldsymbol{x}'^{,\ell-1},
\end{equation}
\begin{equation}
\label{eq:inner2}
\boldsymbol{\hat{x}}^{\ell} = \emph{MLP}(\emph{LN}(\boldsymbol{x}''^{,\ell})) + \boldsymbol{x}''^{,\ell},
\end{equation}
where $\ell \in [1,...,L]$ is the index of the stacked block, $L$ is the number of blocks in Frame Transformer, and $\boldsymbol{\hat{x}}^{\ell} \in \mathbb{R}^{T \times d_f}$ is encoded by the $\ell^{th}$ block. Besides, in Equation (\ref{eq:inner1}) and (\ref{eq:inner2}), $\emph{MSA}(\cdot)$, $\emph{MLP}(\cdot)$ and $\emph{LN}(\cdot)$ are the operations of Multihead Self-Attention (MSA), MultiLayer Perceptron (MLP), and Layer Normalization (LN), respectively, according to \cite{dosovitskiy2020image}, \cite{han2021transformer}. Notably, $\boldsymbol{x}'^{,0}=[\boldsymbol{x}'_1, \boldsymbol{x}'_2,...,\boldsymbol{x}'_T] \in \mathbb{R}^{T \times d_f}$ in Equation (\ref{eq:frame-po}) is the initial input of the frame embedding sequence $\boldsymbol{x}'$.

To aggregate the emotion-related dependencies at different scales, we further design a Segment Transformer to learn frame-level and segment-level correlations of speech emotion. Therefore, the input of Segment Transformer is the combination of the frame-level encoding $\boldsymbol{\hat{x}}$ and segment-level embedding $\boldsymbol{s}$. Specifically, the log-Mel-spectrogram feature $\boldsymbol{x}$ can be divided into a segment set, where each segment $\boldsymbol{s}_j\in\mathbb{R}^{k \times  F \times C}$ consists of $k$ frames, represented as $\boldsymbol{x}=\{\boldsymbol{s}_j\}_{j=1}^{T/k}$. Similar to the Frame Transformer, each segment $\boldsymbol{s}_j$ is firstly transformed to the segment embedding $\boldsymbol{s}'_j \in \mathbb{R}^{1 \times d_s}$ by a linear projection layer $\emph{FC}(\cdot)$ in Segment Transformer. Besides, the $k^{th}$ frame-level encoding $\boldsymbol{\hat{x}}^{s}_j \in \mathbb{R}^{k \times d_f}$ corresponding to the $j^{th}$ segment are also used to aggregate into the segment embeddings after another linear projection $\emph{FC}(\cdot)$, where $\emph{FC}(\cdot)$ is to ensure dimension match for the addition of frame encoding and segment embedding. Then, the combination embedding $\boldsymbol{s}''_j \in \mathbb{R}^{1 \times d_s}$ of frame-level encoding and segment embeddings is generated by
\begin{equation}
\boldsymbol{s}^{\prime}_j = \emph{FC}(\emph{Vec}(\boldsymbol{s}_j)),
\label{eq:input-outer2}
\end{equation}
\begin{equation}
\boldsymbol{\hat{x}}^{s}_j = [\boldsymbol{\hat{x}}_{i+1},\boldsymbol{\hat{x}}_{i+2},..,\boldsymbol{\hat{x}}_{i+k}],
\label{eq:input-outer1}
\end{equation}
\begin{equation}
\boldsymbol{s}''_j = \boldsymbol{s}'_j + \emph{FC}(\emph{Vec}(\boldsymbol{\hat{x}}^{s}_j)),
\label{eq:input-outer3}
\end{equation}
where $\emph{Vec}(\cdot)$ is a vectorization operation to flatten the dimension of $\boldsymbol{\hat{x}}_j^s$ or $\boldsymbol{s}_j$ to $\mathbb{R}^{1 \times (k \times d_f)}$. Then, we also add a learnable class token $\boldsymbol{s}_{\emph{cls}}$ into input sequence for the final emotion classification. Eventually, the segment-level embedding $\boldsymbol{s}'' \in \mathbb{R}^{(T/k+1) \times d_s}$ can be written to
\begin{equation}
\label{eq:segment-emb}
\boldsymbol{s}''=[\boldsymbol{s}_{\emph{cls}}, \boldsymbol{s}''_1, \boldsymbol{s}''_2, ..., \boldsymbol{s}''_{T/k}].
\end{equation}
Similar to Frame Transformer, each segment-level embedding with frame-level aggregation is added the corresponding positions between segments to preserve time-sequence property of inductive bias on speech by a learnable position encoding $\boldsymbol{e}^{s}_{j} \in \mathbb{R}^{1 \times d_s}$, which can be denoted as
\begin{equation}
\boldsymbol{s}''_j \leftarrow \boldsymbol{s}''_j + \boldsymbol{e}^{s}_{j},
\label{eq:segment-po}
\end{equation}
where $\boldsymbol{e}^{s}=\{\boldsymbol{e}^s_j\}_{j=1}^{T/k} \in \mathbb{R}^{(T/k+1) \times d_s}$. The Segment Transformer also adopts $L$ stacked standard transformer blocks to encode the aggregation embedding for the utterance-level representation of speech emotion, where the $\ell^{th}$ block transformations are formalized to
\begin{equation}
\label{eq:outer1}
\boldsymbol{\bar{s}}^{\ell} = \emph{MSA}(\emph{LN}(\boldsymbol{s}''^{,\ell-1})) + \boldsymbol{s}''^{,\ell-1},
\end{equation}
\begin{equation}
\label{eq:outer2}
\boldsymbol{\hat{s}}^{\ell} = \emph{MLP}(\emph{LN}(\boldsymbol{\bar{s}}^{\ell})) + \boldsymbol{\bar{s}}^{\ell},
\end{equation}
where $\boldsymbol{s}''^{,0}$ is the initial segment embedding sequence in Equation (\ref{eq:segment-po}).

With all the above operations, our proposed LGFA firstly models local emotion correlations within frames by Frame Transformer $I(\cdot)$, then aggregates the frame-level encoding $\boldsymbol{\hat{s}}$ and segment embeddings $\boldsymbol{s}''$ to capture global longer-dependencies for the utterance-level emotion representation $\boldsymbol{\hat{s}}$ through Segment Transformer $O(\cdot)$, which can be denoted as
\begin{equation}
\label{eq:fra-seg}
\boldsymbol{\hat{s}} = O(\boldsymbol{s}'';I(\boldsymbol{x}')),
\end{equation}
Furthermore, the class token $\boldsymbol{\hat{s}}_{\emph{cls}}$ can be generated from $\boldsymbol{\hat{s}}$ to input the classifier for speech emotion prediction, represented as
\begin{equation}
\label{eq:classfication}
y_{\emph{pred}}=\emph{C}(\boldsymbol{\hat{s}}_{\emph{cls}}),
\end{equation}
where $y_{\emph{pred}}$, $\emph{C}$, and $\boldsymbol{\hat{s}}_{\emph{cls}}$ are the predicted labels of emotions, classifier, and $\boldsymbol{s}_{\emph{cls}}$ generated by LGFA, respectively. Note that the segment class token $\boldsymbol{s}_{\emph{cls}}$, frame position encoding $\boldsymbol{e}^f$ and segment position encoding $\boldsymbol{e}^s$ are all initialized as zeros in the letter.

\section{Experiments}

In the section, we will introduce the details of our implemented experiments, then discuss the comparison results of the proposed LGFA with state-of-the-art methods.

\emph{Database:} To evaluate the performance of our proposed LGFA, two public emotional speech databases are selected to implement the experiments, i.\,e., the Interactive Emotional Dyadic Motion Capture database (IEMOCAP) \cite{busso2008iemocap} and the China Emotional Database (CASIA) \cite{zhang2008design}. In detail, IEMOCAP is an English multimodal database containing video, speech, and text scripts, which is recorded in 5 sessions (1 male and 1 female in each session) by inducing diverse emotions (\emph{angry, happy, sad, neutral, frustrated, excited, fearful, surprised, disgusted, and others}) of 10 actors under improvised or scripted scenarios. CASIA is a Chinese Emotional Speech Database with 9\,600 recording files under 6 emotions (\emph{angry, fear, happy, neutral, sad, and surprise}). It is collected by inducing 4 actors (2 males and 2 females) to express 6 emotions under several fixed text contents. Note that we adopt 2\,280 improvised samples and 4490 scripted$+$improvised samples with 4 emotions (\emph{angry, happy, sad, and neutral}) in IEMOCAP, and 1\,200 public released samples with 6 emotions in CASIA for experiments.

\emph{Experimental Settings:} In our experiments, all speech sentences are re-sampled to $16$ kHz for Short-Time Fourier Transform (STFT) using 20 ms Hamming window size with 50\% frame overlapping. Then, they are divided into segments with 128 frames as experimental samples and pad 0 for the segment less than 128 frames. Finally, we obtain the log-Mel-spectrogram with the dimension of $\mathbb{R}^{64 \times 128 \times 1}$ for the input of our LGFA, where the number of Mel-filter is set as 64.

For the network of LGFA, the input sizes of Frame Transformer and Segment Transformer are assigned as $(64,128,1)$ and $(64,8,1)$. The number of stacked blocks $L$ is $7$. Furthermore, the projection dimensions and the head number of the Frame Transformer are set as $16$ and $4$, and they are assigned $256$ and $4$ in the Segment Transformer. The LGFA is implemented by PyTorch with NVIDIA A10 GPUs. And it is optimized by the AdamW Optimizer with a learning rate of $0.0001$ and trained from scratch with a batch size of $64$.

In addition, the Leave-One-Subject-Out (LOSO), i.\,e., $k$-fold cross-validation protocol (CV), is adopted for a fair comparison according to \cite{schuller2009acoustic}, \cite{stuhlsatz2011deep}, where $k$ is the speaker number of dataset. Therefore, the speaker rate of training and testing data in IEMOCAP and CASIA are 9:1 and 3:1, respectively. Furthermore, since the IEMOCAP are class-imbalanced, the weighted average recall (WAR) and the unweighted average recall (UAR) \cite{schuller2009acoustic}, \cite{stuhlsatz2011deep} are used to effectively evaluate the performance of the proposed method, where WAR is standard recognition accuracy while UAR is the class-wise accuracy.

\emph{Results and Analysis:} We compare the performance of our proposed LGFA with several state-of-the-art methods on IEMOCAP, i.\,e., CNN$+$LSTM Model \cite{satt2017efficient}, DNN-HMM based model (DNN-HMM\_SGMM-Ali.) \cite{mao2019revisiting}, CNN model with spectrogram (model-2A(spectrogram)) \cite{yenigalla2018speech}, fusion model with different acoustic features (Model-3 (fusion) and Model-1 (dow.$+$ens.)) \cite{bhosale2020deep}. The above methods are all implemented on the improvised data (2280 samples). To further demonstrate the performance of LGFA, we also compare the LGFA with other methods (i.\,e., Bi-LSTM and Greedy$+$Dro.$+$Att.$+$MLP) \cite{huang2016attention} on the scripted$+$improvised data (4490 samples). Moreover, we also choose other comparison methods on CASIA, i.\,e., LLDs with dimension reduction (LLD$+$DR) \cite{liu2018speech}, DNNs with the extreme learning machine (DNN$+$ELM) \cite{han2014speech}, weighted spectral feature learning model (HuWSF) \cite{sun2015weighted}, and DCNN with discriminant temporal pyramid matching (DTPM) \cite{zhang2017speech}. As homologous methods to LGFA, ViT \cite{dosovitskiy2020image} and TNT \cite{han2021transformer} were also used as comparasion methods. Note that the results of DTPM, ViT, and TNT are obtained through our own implementations with the released codes$\footnote{https://github.com/tzaiyang/SpeechEmoRec}^{,}\footnote{https://github.com/lucidrains/vit-pytorch}^{,}\footnote{https://github.com/huawei-noah/CV-Backbones/tree/master/tnt\_pytorch}$.
In addition, to evaluate the experimental performance more comprehensively, these selected comparison methods are based on two commonly used experimental protocols on IEMOCAP, i.\,e., 10-fold LOSO based on speakers and 5-fold LOSO based on sessions. For example, Bi-LSMT, Greedy$+$Dro.$+$Att.$+$MLP, CNN$+$LSTM, DNN-HMM\_SGMM-Ali., ViT, TNT and our proposed LGFA are all based 10-fold CV, other methods are based on 5-fold CV.

\begin{table}[t]
\caption{Experimental results on IEMOCAP, where the best results are highlighted in bold. The first five methods are implemented on the scripted$+$improvised data, and others are based on the improvised data.}
\centering
\renewcommand{\arraystretch}{1.0}

\begin{threeparttable}
\begin{tabular}{p{4.3cm}p{1.3cm}<{\centering}p{1.3cm}<{\centering}p{1.3cm}<{\centering}p{13.cm}<{\centering}}
%\begin{tabular}{ccc}
\toprule          		

\multirow{2}{*}{\bf Comparison Methods}   & \multicolumn{2}{c} {\bf Accuarcy(\%)}  \\ \cline{2-3}
	
                                          & \textbf{WAR} & \textbf{UAR}       \\
				
\midrule
				
Bi-LSTM \cite{huang2016attention}    &  57.87  &  48.54 \cr

%$^{\dag}$Bi-LSTM$+$Att. \cite{huang2016attention}             &  56.38  &  48.70 \cr

Greedy$+$Dro.$+$Att.$+$MLP \cite{huang2016attention}   &  56.33  &  49.96 \cr

ViT                                              &  63.57  &  56.62 \cr

TNT                                              &  63.14  &  56.18 \cr

LGFA (ours)                                          &  \textbf{64.12}  &  \textbf{57.78} \cr

\hline

DNN-HMM\_SGMM-Ali. \cite{mao2019revisiting}          &  62.28  &  58.02 \cr

CNN$+$LSTM \cite{etienne2018cnn+}          &  64.50  &  61.70 \cr

%$^{\ast}$CNN $+$ LSTM \cite{satt2017efficient}                & 68.80   &  59.40 \cr
				
%$^{\ast}$CNN\_GRU-SeqCap \cite{wu2019speech}                  &  72.73  &  59.71 \cr
				
%FCN $+$ Attention \cite{zhang2018attention}          &  70.40  &  63.90 \cr

Model-2A (spectrogram) \cite{yenigalla2018speech}     &  71.30  &  61.60 \cr
				
Model-3 (fusion) \cite{bhosale2020deep}               &  72.34  &  58.31 \cr

%$^{\ast}$Model-1 (dow.$+$ens.) \cite{bhosale2020deep}          &  \textcolor{blue}{70.05}  &  \textcolor{blue}{\textbf{63.27}} \cr

ViT                                              &  70.22  &  58.58 \cr

TNT                                              &  70.61  &  59.72 \cr

%\hline

LGFA (ours)                                          &  \textbf{73.29}  &  \textbf{62.63} \cr

\bottomrule[1.0pt]

\end{tabular}	
\end{threeparttable}
\label{tab:si-iemocap-results} 	
\end{table}

The experimental results with WAR and UAR on IEMOCAP are shown in Table \ref{tab:si-iemocap-results}, where ViT and TNT are implemented by the spectrogram size of 128$\times$128 and the chunk size of 16$\times$16 according to \cite{han2021transformer}, \cite{dosovitskiy2020image}. From these results, it is obvious that the proposed LGFA achieves the competitive performance on both WAR and UAR. Specifically, based on the scripted$+$improvised data, our LGFA improves the accuracies (6.25\% on WAR and 7.82\% on UAR) than comparison methods. Based on the improvised data, LGFA is superior to RNN-based methods (i.\,e., CNN$+$LSTM), demonstrating the advantage of the Transformer-based methods in SER. Further, its results also outperform the ViT and TNT, which reveals LGFA effectively capture the long-range emotion dependencies inside frames and segments for better speech representation and is more suitable for the task of SER than ViT and TNT. Although our LGFA achieve the best performance, the UAR results are lower than the WAR ones on comparison methods because of the class-imbalance in IEMOCAP.

The results on CASIA, illustrated in Table \ref{tab:si-casia-results}, also reveal the superiority of our LGFA (improving $3.17\%$ on WAR and UAR). It is better than traditional methods (i.\,e., LLD$+$DR and HuWSF) and DNN-based approaches (i.\,e., DNN\_ELM and DTPM). Similar to the results on IEMOCAP, our proposed LGFA proves its superiority on the SER again over ViT and TNT. Since CASIA is class-balanced, the results of WAR are equal to those of UAR.

\begin{table}[t]
\caption{Experimental results on CASIA, where the best results are highlighted in bold.}
\centering
\renewcommand{\arraystretch}{1.0}

\begin{threeparttable}
\begin{tabular}{p{4.3cm}p{1.3cm}<{\centering}p{1.3cm}<{\centering}p{1.3cm}<{\centering}p{1.3cm}<{\centering}}
%\begin{tabular}{ccc}
\toprule          		

\multirow{2}{*}{\bf Comparison Methods}   & \multicolumn{2}{c} {\bf Accuarcy(\%)}  \\ \cline{2-3}
	
                                          & \textbf{WAR} & \textbf{UAR}       \\
				
\midrule
				
LLDs$+$DR \cite{liu2018speech}         &  39.50  &  39.50 \cr

DNN$+$ELM                              &  41.17  &  41.17 \cr
				
HuWSF \cite{sun2015weighted}             &  43.50  &  43.50 \cr

DTPM                                 &  45.42  &  45.42 \cr

ViT                                  &  42.83  &  42.83 \cr

TNT                                  &  46.58 &  46.58 \cr

%\hline

LGFA (ours)                          &  \textbf{49.75}  &  \textbf{49.75} \cr

\bottomrule[1.0pt]

\end{tabular}	
\end{threeparttable}
\label{tab:si-casia-results} 	
\end{table}

\begin{figure}[t]
\centering
\includegraphics[width=3.0in,height=2.1in]{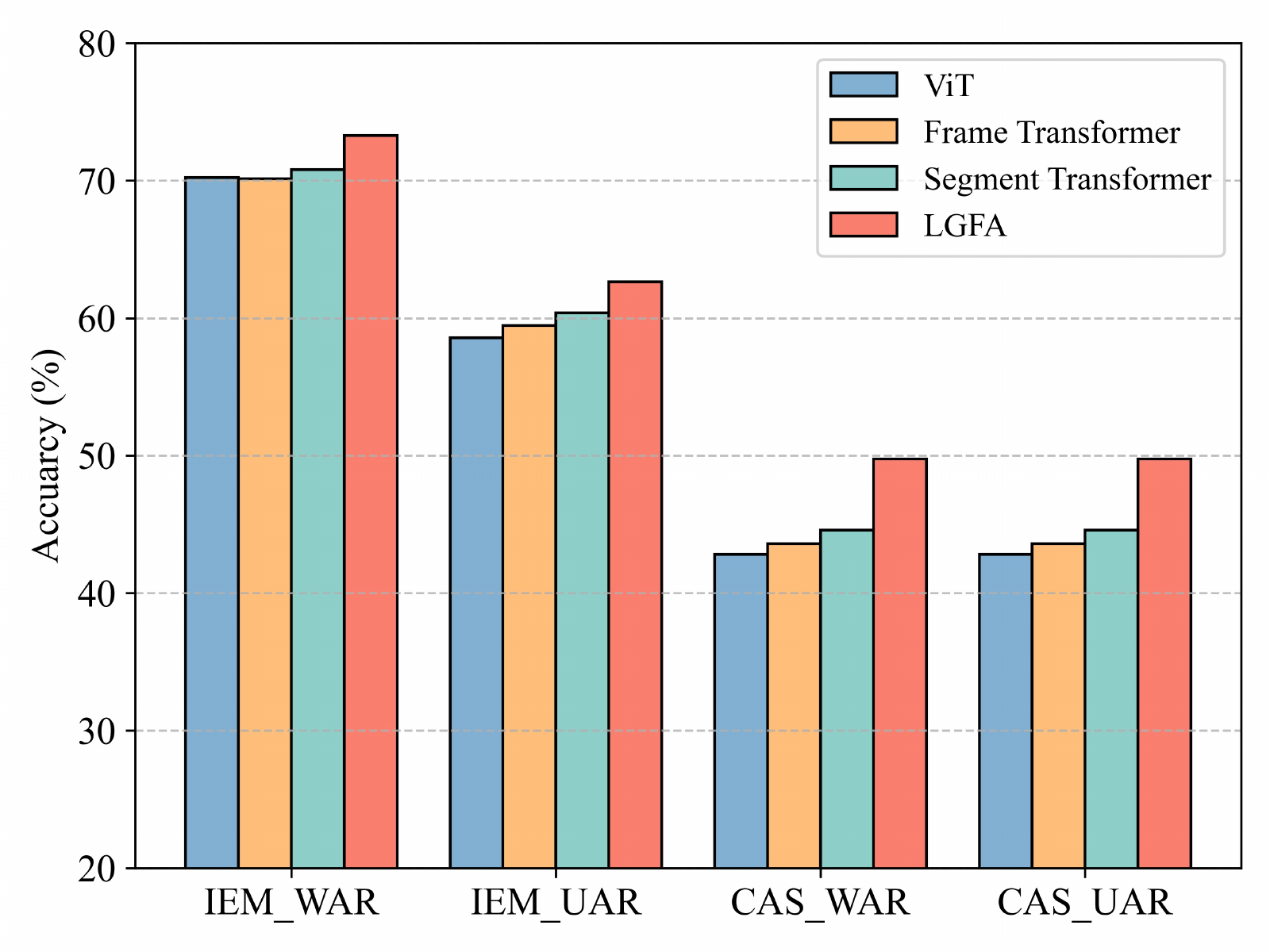}
\caption{Ablation study for LGFA, where IEM\_WAR and CAS\_WAR represent WAR results on IEMOCAP and CASIA, respectively. IEM\_UAR and CAS\_UAR are UAR results.}
\label{fig:ablation}
\end{figure}

Furthermore, to explore the effective components of LGFA, we implement extended experiments to analyze different architectures of our LGFA. Figure. \ref{fig:ablation} shows the results of ablation study, where ViT, Frame Transformer, and Segment Transformer are implemented by square chunks with the size of 16$\times$16, frame chunks with the size of 64$\times$1, segment chunks with the size of 64$\times$8, respectively. The ablation results in Figure. \ref{fig:ablation} indicate that LGFA is superior in speech emotion representation over other architectures. Namely, our designed frame and segment aggregation learning is more suitable for SER than current Speech Transformers. Furthermore, the Segment Transformer outperforms the Frame Transformer, indicating larger chunks will promote the feature extraction of speech emotion for the Transformer.

%\begin{table}[t]
%\caption{Ablation study of LGFA\_T on IEMOCAP and CASIA, where the best results are highlighted in bold.}
%\centering
%\renewcommand{\arraystretch}{1.2}
%
%\begin{threeparttable}
%%\begin{tabular}{p{3cm}p{1cm}<{\centering}p{1cm}<{\centering}p{1cm}<{\centering}p{1cm}<{\centering}}
%\begin{tabular}{ccccc}
%\toprule          		
%
%\multirow{2}{*}{\bf Comparison Methods}   & \multicolumn{2}{c} {\bf IEMOCAP } & \multicolumn{2}{c} {\bf CASIA } \\ \cline{2-5}
%	
%                                          & \textbf{WAR} & \textbf{UAR}       & \textbf{WAR} & \textbf{UAR} \\
%				
%\midrule
%				
%ViT                                   &  70.22        &  58.58            &  42.83        &  42.83             \\
%				
%Frame Transformer                         &  70.13        &  59.47            &  43.58        &  43.58             \\
%
%Segment Transformer                       &  70.79        &  60.38            &  44.58        &  44.58             \\
%
%LGFA                                      &\textbf{73.29} &\textbf{62.63}     &\textbf{49.75} &\textbf{49.75}      \\
%
%
%\bottomrule[1.0pt]
%
%\end{tabular}	
%\end{threeparttable}
%\label{tab:ablation-results} 	
%\end{table}

\emph{Discussion on the extension of LGFA:} In LGFA, to preserve the completeness of the frequency domain in the spectrogram, we divide the spectrogram feature as chunks only on the time domain. To further explore the effect of different chunk division strategies, we extend the chunk division of the proposed LGFA (i.\,e., LGFA\_T in Table \ref{tab:extension-results}) from the time domain to the frequency and time-frequency domain (i.\,e., LGFA\_F and LGFA\_TF in Table \ref{tab:extension-results}). Compared with LGFA\_T, LGFA\_F takes each frequency band as a frame and each frequency band group as a segment to learn the sentence-level emotion feature from the frequency domain. Thus, we can obtain the frequency-wise class token $\boldsymbol{\hat{s}}_{\emph{cls}}^{\emph{fre}}$ of LGFA\_F for the emotions prediction represented as $y_{\emph{pred}}=\emph{C}(\boldsymbol{\hat{s}}_{\emph{cls}}^{\emph{fre}})$. Further, we will also complementarily combine the chunk division methods in the frequency and time domains to generate the fusion class token of LGFA\_TF $\boldsymbol{\hat{s}}_{\emph{cls}}^{fu}=cat(\boldsymbol{\hat{s}}_{\emph{cls}},\boldsymbol{\hat{s}}_{\emph{cls}}^{\emph{fre}})$ to the emotion classifier $y_{\emph{pred}}=\emph{C}(\boldsymbol{\hat{s}}_{\emph{cls}}^{\emph{fu}})$, where $cat(\cdot)$ is the concatenation operation on the feature dimension.

The experimental results of different chunk division strategies are shown in Table \ref{tab:extension-results}. From them, we observe that LGFA\_T and LGFA\_TF outperform LGFA\_F, which may be due to the fact that speech emotion is closely related to the context within frames or segments. While in the frequency domain, not all emotions have obvious energy activations between frequency bands. Furthermore, the LGFA\_TF outperforms LGFA\_T on CASIA, while performs worse on IEMOCAP. The reason may be that chunk division in the frequency domain will not only complement the time-domain chunk division but may also integrate noise caused by the uncertain correlations on the frequency domain under emotions. The recording environment of CASIA contains less noise, while IEMOCAP is recorded in a open dialogue environment. Thus, the noise will affect frequency-domain correlations and impair the performance of the time-frequency fusion model. In other word, the frequency information should be screened to obtain this supplement.

\begin{table}[t]
\caption{Extension experiments of LGFA on the time, frequency, and time-frequency domains, i\,.e., LGFA\_T, LGFA\_F, and LGFA\_TF, where LGFA\_T is the LGFA in Tables \ref{tab:si-iemocap-results} and \ref{tab:si-casia-results}.}
\centering
\renewcommand{\arraystretch}{1.0}

\begin{threeparttable}
%\begin{tabular}{p{3cm}p{1cm}<{\centering}p{1cm}<{\centering}p{1cm}<{\centering}p{1cm}<{\centering}}
\begin{tabular}{ccccc}
\toprule          		

\multirow{2}{*}{\bf Comparison Methods}   & \multicolumn{2}{c} {\bf IEMOCAP } & \multicolumn{2}{c} {\bf CASIA } \\ \cline{2-5}
	
                                          & \textbf{WAR} & \textbf{UAR}       & \textbf{WAR} & \textbf{UAR} \\
				
\midrule
				
LGFA\_T                                   &  \textbf{73.29} &  \textbf{62.63} &  49.75        &  49.75             \\
				
LGFA\_F                                   &  67.85        &  55.57            &  45.33        &  45.33             \\

LGFA\_TF                                  &  70.79        &  60.38            &  \textbf{50.17}  &  \textbf{50.17}             \\

\bottomrule[1.0pt]

\end{tabular}	
\end{threeparttable}
\label{tab:extension-results} 	
\end{table}

\section{Conclusions}

We propose a novel Local to Global Feature Aggregation (LGFA) method for SER. LGFA integrates a Frame Transformer into a Segment Transformer to aggregate local emotion correlations at different scales both within frames and segments for the global utterance-level representation of emotional speech. Through the joint learning of two Transformers, we can obtain discriminative emotion features to learn speech emotion representation from frame-level to segment-level to sentence-level. Extensive experimental results on IEMOCAP and CASIA demonstrate the superiority of our proposed LGFA. Further, we will deeply explore the different chunk division strategies of LGFA for the better SER performance.

\section{Acknowledgements}

%This work was supported in part by the NSFC under Grants U2003207, 61921004, and 61902064, in part by the Jiangsu Frontier Technology Basic Research Project under Grant BK20192004, and in part by the Zhishan Young Scholarship of Southeast University.

This work was supported in part by NSFC under Grant U2003207, in part by National Key R\&D Project under Grant 2022YFC2405600, in part by Jiangsu Frontier Technology Basic Research Project under Grant BK20192004, and in part by Zhishan Young Scholarship of Southeast University.

\bibliographystyle{IEEEtran}
\bibliography{mybib}

% Generated by IEEEtran.bst, version: 1.13 (2008/09/30)
\begin{thebibliography}{10}
\providecommand{\url}[1]{#1}
\csname url@samestyle\endcsname
\providecommand{\newblock}{\relax}
\providecommand{\bibinfo}[2]{#2}
\providecommand{\BIBentrySTDinterwordspacing}{\spaceskip=0pt\relax}
\providecommand{\BIBentryALTinterwordstretchfactor}{4}
\providecommand{\BIBentryALTinterwordspacing}{\spaceskip=\fontdimen2\font plus
\BIBentryALTinterwordstretchfactor\fontdimen3\font minus
  \fontdimen4\font\relax}
\providecommand{\BIBforeignlanguage}[2]{{%
\expandafter\ifx\csname l@#1\endcsname\relax
\typeout{** WARNING: IEEEtran.bst: No hyphenation pattern has been}%
\typeout{** loaded for the language `#1'. Using the pattern for}%
\typeout{** the default language instead.}%
\else
\language=\csname l@#1\endcsname
\fi
#2}}
\providecommand{\BIBdecl}{\relax}
\BIBdecl

\bibitem{schuller2021intelligent}
B.~W. Schuller, R.~Picard, E.~Andr{\'e}, J.~Gratch, and J.~Tao, ``Intelligent
  signal processing for affective computing,'' 2021.

\bibitem{cowie2001emotion}
R.~Cowie, E.~Douglas-Cowie, N.~Tsapatsoulis, G.~Votsis, S.~Kollias, W.~Fellenz,
  and J.~G. Taylor, ``Emotion recognition in human-computer interaction,''
  \emph{IEEE Signal Processing Magazine}, vol.~18, no.~1, pp. 32--80, 2001.

\bibitem{schuller2009acoustic}
B.~Schuller, B.~Vlasenko, F.~Eyben, G.~Rigoll, and A.~Wendemuth, ``Acoustic
  emotion recognition: A benchmark comparison of performances,'' in \emph{2009
  IEEE Workshop on Automatic Speech Recognition \& Understanding}.\hskip 1em
  plus 0.5em minus 0.4em\relax IEEE, 2009, pp. 552--557.

\bibitem{stuhlsatz2011deep}
A.~Stuhlsatz, C.~Meyer, F.~Eyben, T.~Zielke, G.~Meier, and B.~Schuller, ``Deep
  neural networks for acoustic emotion recognition: Raising the benchmarks,''
  in \emph{2011 IEEE International Conference on Acoustics, Speech and Signal
  Processing (ICASSP)}.\hskip 1em plus 0.5em minus 0.4em\relax IEEE, 2011, pp.
  5688--5691.

\bibitem{lu2022domain}
C.~Lu, Y.~Zong, W.~Zheng, Y.~Li, C.~Tang, and B.~W. Schuller, ``Domain
  invariant feature learning for speaker-independent speech emotion
  recognition,'' \emph{IEEE/ACM Transactions on Audio, Speech, and Language
  Processing}, vol.~30, pp. 2217--2230, 2022.

\bibitem{chen20183}
M.~Chen, X.~He, J.~Yang, and H.~Zhang, ``3-D convolutional recurrent neural
  networks with attention model for speech emotion recognition,'' \emph{IEEE
  Signal Processing Letters}, vol.~25, no.~10, pp. 1440--1444, 2018.

\bibitem{zhang2008design}
J.~T. F. L.~M. Zhang and H.~Jia, ``Design of speech corpus for mandarin text to
  speech,'' in \emph{The Blizzard Challenge 2008 workshop}, 2008.

\bibitem{satt2017efficient}
A.~Satt, S.~Rozenberg, and R.~Hoory, ``Efficient emotion recognition from
  speech using deep learning on spectrograms.'' in \emph{INTERSPEECH}, 2017,
  pp. 1089--1093.

\bibitem{wu2019speech}
X.~Wu, S.~Liu, Y.~Cao, X.~Li, J.~Yu, D.~Dai, X.~Ma, S.~Hu, Z.~Wu, X.~Liu
  \emph{et~al.}, ``Speech emotion recognition using capsule networks,'' in
  \emph{ICASSP 2019-2019 IEEE International Conference on Acoustics, Speech and
  Signal Processing (ICASSP)}.\hskip 1em plus 0.5em minus 0.4em\relax IEEE,
  2019, pp. 6695--6699.

\bibitem{yu2019review}
Y.~Yu, X.~Si, C.~Hu, and J.~Zhang, ``A review of recurrent neural networks:
  LSTM cells and network architectures,'' \emph{Neural Computation}, vol.~31,
  no.~7, pp. 1235--1270, 2019.

\bibitem{vaswani2017attention}
A.~Vaswani, N.~Shazeer, N.~Parmar, J.~Uszkoreit, L.~Jones, A.~N. Gomez,
  {\L}.~Kaiser, and I.~Polosukhin, ``Attention is all you need,''
  \emph{Advances in Neural Information Processing Systems}, vol.~30, 2017.

\bibitem{tarantino2019self}
L.~Tarantino, P.~N. Garner, A.~Lazaridis \emph{et~al.}, ``Self-attention for
  speech emotion recognition.'' in \emph{INTERSPEECH}, 2019, pp. 2578--2582.

\bibitem{gong2021ast}
Y.~Gong, Y.-A. Chung, and J.~Glass, ``AST: Audio Spectrogram Transformer,''
  \emph{arXiv preprint arXiv:2104.01778}, 2021.

\bibitem{dosovitskiy2020image}
A.~Dosovitskiy, L.~Beyer, A.~Kolesnikov, D.~Weissenborn, X.~Zhai,
  T.~Unterthiner, M.~Dehghani, M.~Minderer, G.~Heigold, S.~Gelly \emph{et~al.},
  ``An image is worth 16x16 words: Transformers for image recognition at
  scale,'' \emph{arXiv preprint arXiv:2010.11929}, 2020.

\bibitem{lin2021chunk}
W.-C. Lin and C.~Busso, ``Chunk-level speech emotion recognition: A general
  framework of sequence-to-one dynamic temporal modeling,'' \emph{IEEE
  Transactions on Affective Computing}, 2021.

\bibitem{zhang2017speech}
S.~Zhang, S.~Zhang, T.~Huang, and W.~Gao, ``Speech emotion recognition using
  deep convolutional neural network and discriminant temporal pyramid
  matching,'' \emph{IEEE Transactions on Multimedia}, vol.~20, no.~6, pp.
  1576--1590, 2017.

\bibitem{shen2020wise}
G.~Shen, R.~Lai, R.~Chen, Y.~Zhang, K.~Zhang, Q.~Han, and H.~Song, ``WISE:
  Word-level interaction-based multimodal fusion for speech emotion
  recognition.'' in \emph{INTERSPEECH}, 2020, pp. 369--373.

\bibitem{han2021transformer}
K.~Han, A.~Xiao, E.~Wu, J.~Guo, C.~Xu, and Y.~Wang, ``Transformer in
  Transformer,'' \emph{Advances in Neural Information Processing Systems},
  vol.~34, 2021.

\bibitem{busso2008iemocap}
C.~Busso, M.~Bulut, C.-C. Lee, A.~Kazemzadeh, E.~Mower, S.~Kim, J.~N. Chang,
  S.~Lee, and S.~S. Narayanan, ``IEMOCAP: Interactive emotional dyadic motion
  capture database,'' \emph{Language Resources and Evaluation}, vol.~42, no.~4,
  pp. 335--359, 2008.

\bibitem{mao2019revisiting}
S.~Mao, D.~Tao, G.~Zhang, P.~Ching, and T.~Lee, ``Revisiting hidden markov
  models for speech emotion recognition,'' in \emph{ICASSP 2019-2019 IEEE
  International Conference on Acoustics, Speech and Signal Processing
  (ICASSP)}.\hskip 1em plus 0.5em minus 0.4em\relax IEEE, 2019, pp. 6715--6719.

\bibitem{yenigalla2018speech}
P.~Yenigalla, A.~Kumar, S.~Tripathi, C.~Singh, S.~Kar, and J.~Vepa, ``Speech
  emotion recognition using spectrogram \& phoneme embedding.'' in
  \emph{INTERSPEECH}, 2018, pp. 3688--3692.

\bibitem{bhosale2020deep}
S.~Bhosale, R.~Chakraborty, and S.~K. Kopparapu, ``Deep encoded linguistic and
  acoustic cues for attention based end to end speech emotion recognition,'' in
  \emph{ICASSP 2020-2020 IEEE International Conference on Acoustics, Speech and
  Signal Processing (ICASSP)}.\hskip 1em plus 0.5em minus 0.4em\relax IEEE,
  2020, pp. 7189--7193.

\bibitem{huang2016attention}
C.-W. Huang and S.~S. Narayanan, ``Attention assisted discovery of
  sub-utterance structure in speech emotion recognition.'' in
  \emph{INTERSPEECH}, 2016, pp. 1387--1391.

\bibitem{liu2018speech}
Z.-T. Liu, Q.~Xie, M.~Wu, W.-H. Cao, Y.~Mei, and J.-W. Mao, ``Speech emotion
  recognition based on an improved brain emotion learning model,''
  \emph{Neurocomputing}, vol. 309, pp. 145--156, 2018.

\bibitem{han2014speech}
K.~Han, D.~Yu, and I.~Tashev, ``Speech emotion recognition using deep neural
  network and extreme learning machine,'' in \emph{INTERSPEECH 2014}, 2014.

\bibitem{sun2015weighted}
Y.~Sun, G.~Wen, and J.~Wang, ``Weighted spectral features based on local Hu
  moments for speech emotion recognition,'' \emph{Biomedical Signal Processing
  and Control}, vol.~18, pp. 80--90, 2015.

\bibitem{etienne2018cnn+}
C.~Etienne, G.~Fidanza, A.~Petrovskii, L.~Devillers, and B.~Schmauch, ``CNN +
  LSTM architecture for speech emotion recognition with data augmentation,'' in
  \emph{Workshop on Speech, Music and Mind 2018}.\hskip 1em plus 0.5em minus
  0.4em\relax ISCA, 2018, pp. 21--25.

\end{thebibliography}

\end{document}